\documentclass[aps,prd,twocolumn,superscriptaddress,nofootinbib,10pt]{revtex4-2}

\usepackage{amsmath,amssymb,mathrsfs,bm}
\usepackage{graphicx}
\usepackage{xcolor}
\usepackage{amsthm}
\usepackage[colorlinks=true,linkcolor=blue,citecolor=blue,urlcolor=blue]{hyperref}
\usepackage{orcidlink}

\newcommand{\dd}{\mathrm{d}}
\newcommand{\ee}{\mathrm{e}}
\newcommand{\cR}{{}^{(3)}\!R}
\newcommand{\eps}{\varepsilon}

\newcommand{\Tr}{\operatorname{Tr}}
\newcommand{\mathscrQ}{\mathcal Q}

\begin{document}

\title{Intrinsic spatial-strain shortcuts in general relativity}

\author{Francisco S. N. Lobo\,\orcidlink{0000-0002-9388-8373}}
\email{fslobo@ciencias.ulisboa.pt}
\affiliation{Instituto de Astrof\'isica e Ci\^encias do Espa\c{c}o, Faculdade de Ci\^encias da Universidade de Lisboa, Edif\'icio C8, Campo Grande, P-1749-016 Lisboa, Portugal}
\affiliation{Departamento de F\'isica, Faculdade de Ci\^encias da Universidade de Lisboa, Edif\'icio C8, Campo Grande, P-1749-016 Lisboa, Portugal}

\date{\today}

\begin{abstract}
We analyze spacetime shortcuts generated by longitudinal contraction and transverse expansion of the intrinsic spatial metric at fixed reference volume. Smooth bounded ultrastatic corridors are globally hyperbolic and geodesically complete and produce an exact travel-time advance between exterior clocks. An axial maximum makes the axial light ray globally minimizing; a strict maximum on an interior interval establishes Olum's endpoint-surface criterion for exactly matched corridors. Exact Einstein-source identities distinguish integrated Eulerian matter energy from asymptotic mass: infinite tubes have negative integrated energy, whereas finite corridors admit competing gradient contributions. A transverse Ricci identity proves local null-energy violation in every nontrivial localized finite corridor. The axial averaged null energy is strictly negative for localized axial-maximum profiles, and quantitative bounds relate its magnitude to the reference time advance and peak strain. We also obtain the complete finite-profile Ricci spectrum and the axial passenger tidal tensor. These results provide explicit causal and source constraints on a prepared geometry without specifying its physical formation or matter dynamics.
\end{abstract}

\maketitle

\section{Introduction}
\label{sec:introduction}

General relativity permits several distinct geometrical mechanisms for reducing travel time relative to a specified reference spacetime. Traversable wormholes change global connectivity~\cite{MorrisThorne1988,MorrisThorneYurtsever1988}; Alcubierre and Nat\'ario geometries place their principal transport mechanism in a nontrivial shift on flat spatial slices~\cite{Alcubierre1994,Natario2002}; and Krasnikov tubes alter the causal geometry of a prepared route~\cite{Krasnikov1998,EverettRoman1997,Krasnikov2003}. Coordinate speed alone does not establish a physical shortcut. A comparison requires specified endpoints, clock normalization, and causal competitors~\cite{Olum1998,LoboVisser2004}. We use a boundary-matched Minkowski reference and prove Olum's endpoint-surface condition for a complete ultrastatic subclass with a strict axial maximum and exactly flat endpoint neighborhoods. Reviews of exotic geometries and their source requirements provide the broader setting for this construction~\cite{LoboReview2008,LoboBook2017}. The geometric use of a prepared route, local causal propagation and the separate formation problem are discussed in~\cite{AlcubierreLobo2017}.

The supporting source is a central obstacle. Quantum inequalities constrain conventional thin-wall warp constructions~\cite{PfenningFord1997}, while weak-field analyses show that their energy-condition violations persist at low velocities~\cite{LoboVisser2004}. Van Den Broeck demonstrated that changes to spatial geometry can alter volume and integrated source requirements~\cite{VanDenBroeck1999}. Recent proposals extend the range of metric ans\"atze~\cite{BobrickMartire2021,Lentz2021}, and numerical tools permit systematic source assessment~\cite{Helmerich2024}. However, positive density for Eulerian observers does not establish the weak energy condition (WEC), which requires nonnegative energy density for every timelike observer; analogous care is required for the null energy condition (NEC)~\cite{SantiagoSchusterVisser2022}. Broader treatments restore spatial curvature and covariant kinematics~\cite{BarzegarBuchert2025,Chowdhury2025,BarzegarBuchertVigneron2026,BuchertFrackowiak2026}. Unit-determinant nonflat spatial deformations also occur in unit-lapse representations of Kerr~\cite{VisserLiberati2022}.

We isolate a specific intrinsic mechanism: longitudinal proper distances contract while the two transverse directions expand, preserving the volume form relative to a fixed Euclidean reference metric. The static subclass already yields a travel-time advance, so a nonzero shift is unnecessary. The contribution is the exact causal and source analysis of this strain family, rather than a priority claim for spatial curvature or volume preservation themselves. Smooth localization creates curvature in the transition regions; a constant strain removable by a coordinate rescaling is not the construction of interest.

The distinction between local, averaged, and volume-integrated source measures is especially useful here. A finite static corridor admits competing radial and longitudinal contributions to its integrated Eulerian energy. Nevertheless, zero-ADM-mass rigidity forces negative local energy somewhere in every nonflat, complete, asymptotically Euclidean member. For profiles maximal on the axis, an independent axial averaged-null-energy identity gives a causal source obstruction and quantitative advance bounds. A second, transverse Ricci identity establishes NEC violation in localized finite corridors without an axial-maximum hypothesis. The full Ricci spectrum and passenger-frame tides provide complementary local indicators. These results complement general time-delay theorems, whose energy and genericity hypotheses must be checked rather than assumed~\cite{GaoWald2000}.

The paper is organized as follows:
Section~\ref{sec:geometry} defines the reference-relative strain and its global causal properties. Section~\ref{sec:travel} establishes the endpoint comparison, global minimality, the strict endpoint-surface criterion, an explicit matching construction and the optical interpretation. Section~\ref{sec:sources} derives the full static Ricci spectrum, source identities, local and averaged energy obstructions, and quantitative advance bounds. Section~\ref{sec:physical} discusses their physical scope, optimization, and tensorial extensions, and Sec.~\ref{sec:conclusions} summarizes the results and outlines future research.

\section{Geometry, regularity, and causality}
\label{sec:geometry}

Throughout we use geometrized units, $G=c=1$, and metric signature $(-+++)$. In the Arnowitt--Deser--Misner (ADM) decomposition, the spacetime metric is written as
\begin{equation}
	\dd s^2=-\alpha^2\dd t^2+
	\gamma_{ij}(\dd x^i+\beta^i\dd t)(\dd x^j+\beta^j\dd t),
	\label{eq:adm}
\end{equation}
Here $\alpha$ is the lapse, $\beta^i$ is the shift, and $\gamma_{ij}$ is the induced spatial metric on a slice of constant $t$. The lapse relates coordinate time to proper time for observers normal to the slices, while the shift describes the displacement of spatial coordinates between slices. Greek indices denote spacetime components, and Latin indices denote spatial components. We use the extrinsic-curvature convention
\begin{equation}
	K_{ij}=\frac{1}{2\alpha}
	\left(D_i\beta_j+D_j\beta_i-\partial_t\gamma_{ij}\right).
	\label{eq:Kconv}
\end{equation}

The derivative $D_i$ is the covariant derivative compatible with $\gamma_{ij}$, and $\beta_i=\gamma_{ij}\beta^j$. Spatial indices are raised and lowered with $\gamma_{ij}$. A comma denotes an ordinary coordinate derivative, for example $\Psi_{,z}=\partial_z\Psi$ and $\Psi_{,\varpi z}=\partial_\varpi\partial_z\Psi$.

\subsection{Spacetime metric}

Let $(t,\varpi,\phi,z)$ be cylindrical coordinates, with $\varpi\ge0$ the cylindrical radius, $\phi$ the azimuthal angle, and $z$ the longitudinal coordinate along the corridor. Consider
\begin{align}
\dd s^2={}&-\alpha^2\dd t^2+\ee^{\Psi}\left(\dd\varpi^2+\varpi^2\dd\phi^2\right)\nonumber\\
&+\ee^{-2\Psi}\left(\dd z-V\dd t\right)^2.
\label{eq:mastermetric}
\end{align}

Here $\alpha>0$, $\Psi$, and $V$ are smooth axisymmetric functions of $(t,\varpi,z)$. The dimensionless scalar $\Psi$ controls the intrinsic spatial strain, and $V$ is the axial shift function, with units of velocity before setting $c=1$. All explicit source calculations below are stationary and use $\alpha=1$. The subclass $V=0$, $\alpha=1$ is ultrastatic: its time direction is orthogonal to time-independent spatial slices, and $t$ is proper time for static observers. The spatial metric and shift are
\begin{align}
\gamma_{ij}\dd x^i\dd x^j
&=\ee^{\Psi}(\dd\varpi^2+\varpi^2\dd\phi^2)
+\ee^{-2\Psi}\dd z^2,
\label{eq:spatialmetric}\\
\beta^i\partial_i&=-V\partial_z.
\label{eq:shift}
\end{align}
To formulate the strain invariantly, introduce the Euclidean reference three-metric
\begin{equation}
\bar\gamma_{ij}\dd x^i\dd x^j
=\dd\varpi^2+\varpi^2\dd\phi^2+\dd z^2
\label{eq:reference3}
\end{equation}
on the same manifold and define the mixed deformation tensor
\begin{equation}
C^i{}_j\equiv\bar\gamma^{ik}\gamma_{kj}.
\label{eq:Ctensor}
\end{equation}
Its principal eigenvalues are
\begin{equation}
\lambda_\varpi=\ee^{\Psi},\qquad
\lambda_\phi=\ee^{\Psi},\qquad
\lambda_z=\ee^{-2\Psi},
\end{equation}
and therefore $\det C=1$.

These eigenvalues are squared proper-length factors relative to the reference metric. They should be distinguished from the Ricci eigenvalues introduced in Sec.~\ref{sec:Riccispectrum}, which have units of inverse length squared.

Equivalently, the physical and reference volume forms coincide,
\begin{equation}
{\rm vol}_{\gamma}={\rm vol}_{\bar\gamma}
=\varpi\,\dd\varpi\wedge\dd\phi\wedge\dd z,
\label{eq:volumeforms}
\end{equation}
which in the chosen cylindrical chart is represented by
\begin{equation}
\det\gamma=\varpi^2,\qquad \sqrt{\gamma}=\varpi.
\label{eq:detgamma}
\end{equation}
Thus ``volume preserving'' means the coordinate-independent relation \(\det(\bar\gamma^{-1}\gamma)=1\), not equality of bare determinants in arbitrary coordinates.  Longitudinal proper lengths are multiplied by \(\ee^{-\Psi}\), while each transverse proper length is multiplied by \(\ee^{\Psi/2}\).

This reference-relative unit-determinant property is the defining feature of the present strain ansatz.  It differs from a conformal rescaling such as the spatial factor used in the Van Den Broeck geometry \cite{VanDenBroeck1999}, where the local spatial volume is deliberately enlarged.  It also differs from Nat\'ario's zero-expansion condition \cite{Natario2002}: there, volume preservation is a kinematical condition on a shift flow in a flat spatial metric; here it is built directly into the intrinsic spatial geometry.

For \(\alpha=1\), the nonzero metric components in the \((t,z)\) block are
\begin{equation}
g_{tt}=-1+\ee^{-2\Psi}V^2,
\quad g_{tz}=-\ee^{-2\Psi}V,
\quad g_{zz}=\ee^{-2\Psi},
\end{equation}
and the inverse block is
\begin{equation}
g^{tt}=-1,
\quad g^{tz}=-V,
\quad g^{zz}=\ee^{2\Psi}-V^2.
\label{eq:inverse}
\end{equation}
The full determinant is $\det g=-\alpha^2\varpi^2$,
independent of \(\Psi\) and \(V\).  The zero at \(\varpi=0\) is the usual coordinate degeneracy of cylindrical coordinates, not a metric degeneracy.

Regularity must be assessed in nonsingular Cartesian coordinates near the symmetry axis.  A convenient sufficient condition is that \(\Psi\), \(V\), and \(\alpha\) extend as \(C^2\) axisymmetric functions of \(x^2+y^2=\varpi^2\) and \(z\).  In particular,
\begin{equation}
\partial_\varpi\Psi|_{\varpi=0}
=\partial_\varpi V|_{\varpi=0}
=\partial_\varpi\alpha|_{\varpi=0}=0,
\label{eq:axisreg}
\end{equation}
with finite \(\Psi,V,\alpha\) and \(\alpha>0\).  Under these assumptions the apparent zero of \(\det g\) at \(\varpi=0\) is only the standard cylindrical-coordinate degeneracy; the metric and inverse metric are \(C^2\) in regular Cartesian charts, and polynomial curvature invariants are finite for finite strain.  The determinant check alone is not a regularity proof.

\subsection{Causality, completeness, and geometrical nontriviality}

For the unit-lapse metric, Eq.~\eqref{eq:inverse} gives
\begin{equation}
g^{\mu\nu}(\nabla_\mu t)(\nabla_\nu t)=g^{tt}=-1.
\label{eq:timefunction}
\end{equation}

Thus $t$ is a temporal function on the global product $\mathbb R\times\mathbb R^3$, with future orientation chosen by increasing $t$, and the spacetime is stably causal. In fact $g^{tt}=-\alpha^{-2}<0$ also holds for arbitrary positive lapse. An ADM chart alone does not establish global hyperbolicity~\cite{BarzegarBuchertVigneron2026}.

For the ultrastatic subclass, $V=0$, $\alpha=1$, let $\Psi$ be smooth and uniformly bounded on $\mathbb R^3$. The Cartesian spatial metric obeys $c_-\bar\gamma\le\gamma\le c_+\bar\gamma$ for positive constants $c_\pm$. It is consequently complete. Along every future causal curve its spatial arc length is bounded by the elapsed $t$. If an inextendible causal curve had a finite upper limit of $t$, completeness would give a spatial endpoint and hence a spacetime extension, a contradiction. The same argument applies to the past, so each $t=\mathrm{const}$ slice is Cauchy and the product is globally hyperbolic~\cite{ONeill1983}. No compact-support assumption is needed for this argument.

The constants $c_-$ and $c_+$ bound physical squared lengths relative to Euclidean squared lengths. A Cauchy slice intersects every inextendible causal curve exactly once; global hyperbolicity therefore gives a well-defined causal initial-value setting for the prescribed geometry.

The ultrastatic geodesic equations further give $\dd^2t/\dd\lambda^2=0$ and spatial geodesic motion with constant speed. Completeness of the spatial metric therefore implies spacetime geodesic completeness. For compactly supported or sufficiently rapidly decaying strain there is no event horizon: every point can send a causal signal into the asymptotically Minkowski exterior and onward to future null infinity. These global conclusions apply to the stated ultrastatic subclass, not automatically to arbitrary moving or shifted extensions.

It is also important to separate a physical strain field from a coordinate artifact.  If \(\Psi=\Psi_0\) is constant everywhere and \(V=0\), then the transformations
\begin{equation}
\tilde\varpi=\ee^{\Psi_0/2}\varpi,
\qquad
\tilde z=\ee^{-\Psi_0}z
\end{equation}
make the metric Minkowskian.  Therefore a constant strain is not a shortcut by itself.  The geometry becomes nontrivial only because \(\Psi\to0\) in the exterior and hence possesses transition regions where \(\nabla_i\Psi\neq0\).  Curvature is concentrated in those walls and endcaps.  This observation is the spatial-strain analogue of the familiar fact that the flat passenger region of an Alcubierre bubble does not remove the curvature of its wall.

\section{Travel-time advance and optical geometry}
\label{sec:travel}

\subsection{Local light cones}

Set \(\dd\varpi=\dd\phi=0\).  The axial null condition from Eq.~\eqref{eq:mastermetric} gives
\begin{equation}
\frac{\dd z}{\dd t}=V\pm\alpha\ee^{\Psi}.
\label{eq:nullspeed}
\end{equation}
For \(\alpha=1\), a static strain field with \(V=0\) already gives
\begin{equation}
\left|\frac{\dd z}{\dd t}\right|=\ee^{\Psi}.
\end{equation}
This coordinate speed has no local superluminal meaning by itself.  Its physical content follows only after a reference geometry and endpoint normalization are specified.  We take the physical spacetime to agree with Minkowski spacetime in neighborhoods of two endpoint worldlines, so that \(t\) and \(z\) there have the same normalization in the physical and reference geometries.

The Eulerian unit normal is
\begin{equation}
n^\mu=\alpha^{-1}\left(1,0,0,V\right),
\end{equation}
and the unit axial spatial vector is
$e_{\hat z}=\ee^{\Psi}\partial_z$.
A massive traveler with local Eulerian velocity \(u\), \(|u|<1\), has
\begin{equation}
U^\mu=\Gamma\left(n^\mu+u e_{\hat z}^\mu\right),
\qquad \Gamma=(1-u^2)^{-1/2},
\end{equation}
so that
\begin{equation}
\frac{\dd z}{\dd t}=V+\alpha u\ee^{\Psi}.
\label{eq:massivespeed}
\end{equation}

For the ultrastatic subclass, the endpoint comparison below can give an arrival earlier than the reference-Minkowski light signal while $|u|<1$. This statement compares two specified geometries; every physical trajectory remains inside the local physical light cone.

\subsection{Boundary-matched reference time advance}

Let $A$ and $B$ be static endpoint worldlines on the symmetry axis, with $z_A<z_B$. For the exact boundary-matched construction, take $\Psi=0$ in neighborhoods of both endpoints and use the global ultrastatic time to synchronize their clocks. Introduce the reference spacetime
\begin{equation}
\dd\bar s^2=-\dd t^2+\dd\varpi^2+\varpi^2\dd\phi^2+\dd z^2,
\label{eq:ref4}
\end{equation}
identified with the physical spacetime in those endpoint neighborhoods.  Let \(L=z_B-z_A\) denote the corresponding reference axial separation.

For the static unit-lapse subclass, $V=0$ and $\alpha=1$,
smooth axisymmetry implies that the symmetry axis is a totally geodesic submanifold: it is the fixed-point set of the rotational isometry.  Hence an axial null curve is an actual null geodesic.  Its travel time is
\begin{equation}
T_{\rm phys}=\int_{z_A}^{z_B}\ee^{-\Psi(0,z)}\,\dd z.
\label{eq:Taxis}
\end{equation}
The corresponding reference-Minkowski null travel time is
$T_{\rm ref}=L$. We define the boundary-matched reference time advance
\begin{equation}
\Delta T_{\rm ref}\equiv
T_{\rm ref}-T_{\rm phys}
=L-\int_{z_A}^{z_B}\ee^{-\Psi(0,z)}\,\dd z.
\label{eq:deltaTref}
\end{equation}

\textit{Time-advance theorem.}  If \(\Psi(0,z)\ge0\) along the route and is positive on a set of nonzero measure, then
\begin{equation}
\Delta T_{\rm ref}>0.
\label{eq:shortcutineq}
\end{equation}
The proof follows directly from \(\ee^{-\Psi}\le1\), with strict inequality where \(\Psi>0\).

A freely moving material traveler following the axial spatial geodesic with constant local Eulerian speed \(0<u<1\) has
\begin{equation}
T_{\rm ship}=\frac{1}{u}\int_{z_A}^{z_B}\ee^{-\Psi(0,z)}\,\dd z,
\label{eq:Tship}
\end{equation}
so \(T_{\rm ship}<T_{\rm ref}\) whenever
\begin{equation}
u>\frac{1}{L}\int_{z_A}^{z_B}\ee^{-\Psi(0,z)}\,\dd z.
\label{eq:timelikecriterion}
\end{equation}

If most of the reference endpoint separation lies in a plateau $\Psi\simeq\Psi_0$, with negligible undeformed segments and transition lengths, this reduces to $u\gtrsim\ee^{-\Psi_0}$. Otherwise Eq.~\eqref{eq:timelikecriterion}, including the entire route, is required. The traveler proper time is $\tau_{\rm ship}=T_{\rm ship}\sqrt{1-u^2}$; the time advance itself uses the synchronized endpoint clocks.

A useful sufficient condition also determines the fastest physical path. Suppose
\begin{equation}
\Psi(\varpi,z)\le\psi(z),\qquad \psi(z)\equiv\Psi(0,z).
\label{eq:axismaximum}
\end{equation}
For any piecewise smooth spatial competitor $x(s)$ joining the axial endpoints, where $s$ is an arbitrary curve parameter and an overdot denotes $\dd/\dd s$,
\begin{align}
\ell_\gamma[x]
&=\int\!\sqrt{\ee^\Psi(\dot\varpi^2+\varpi^2\dot\phi^2)+\ee^{-2\Psi}\dot z^2}\,\dd s\nonumber\\
&\ge\int\!\ee^{-\psi(z)}|\dot z|\,\dd s
\ge\int_{z_A}^{z_B}\ee^{-\psi(z)}\,\dd z.
\label{eq:globalminimum}
\end{align}
The axis saturates both inequalities. Thus it globally minimizes spatial optical length and null travel time between these axial endpoints; the argument allows off-axis paths and longitudinal backtracking. Applied to every pair of points on the complete axial null ray, it shows that ray is achronal.

Here achronal means that no two points on the ray can be joined by a timelike curve. The weak maximum condition proves point-to-point minimality. The strict condition below additionally controls all pairs of points on endpoint two-surfaces~\cite{Olum1998}.

\subsection{Strict endpoint-surface minimality}
\label{sec:Olum}

The reference advance is a comparison between specified geometries. An independent invariant statement follows from a strengthened maximum assumption. Let $V=0$, $\alpha=1$, and let the spatial metric be smooth, complete and exactly Euclidean in neighborhoods of the endpoint events. In addition to Eq.~\eqref{eq:axismaximum}, suppose there is a nonempty open interval $J\subset(z_A,z_B)$ such that
\begin{equation}
 \Psi(\varpi,z)<\psi(z)\quad(\varpi>0,\ z\in J).
 \label{eq:strictaxis}
\end{equation}
Let $\mathsf A=(0,0,0,z_A)$ and $\mathsf B=(T_{\rm phys},0,0,z_B)$ in Cartesian coordinates $(t,x,y,z)$. Choose sufficiently small disks
\begin{align}
 \Sigma_A&=\{t=0,\ z=z_A,\ x^2+y^2<d_A^2\},\nonumber\\
 \Sigma_B&=\{t=T_{\rm phys},\ z=z_B,\ x^2+y^2<d_B^2\}
 \label{eq:endpointdisks}
\end{align}
contained in the flat neighborhoods.

The positive constants $d_A$ and $d_B$ are the disk radii. The notation $J^+(p)$ denotes the causal future of the event $p$, including points reached by null curves; it is distinct from the longitudinal interval $J$ in Eq.~\eqref{eq:strictaxis}. Every point of each disk is connected to its center by a spacelike geodesic lying in that disk. We claim
\begin{equation}
 q\in J^+(p),\quad p\in\Sigma_A,\ q\in\Sigma_B
 \quad\Longleftrightarrow\quad(p,q)=(\mathsf A,\mathsf B).
 \label{eq:Olumcondition}
\end{equation}
To prove this, the lower bound~\eqref{eq:globalminimum} applies to any spatial path between the disks, since its endpoint $z$ coordinates are still $z_A,z_B$. A path saturating the bound must have zero Cartesian transverse velocity and monotone longitudinal motion. Its transverse position is constant. If this position is nonzero, Eq.~\eqref{eq:strictaxis} makes the length strictly larger on $J$. Thus only the axis can saturate the bound. By spatial completeness, Hopf--Rinow guarantees a minimizing geodesic between each pair of spatial endpoints~\cite{ONeill1983}. Consequently their distance is strictly greater than $T_{\rm phys}$ unless both endpoints are the centers. In an ultrastatic product, causal relatedness is equivalent to elapsed time being at least spatial distance. This proves Eq.~\eqref{eq:Olumcondition}, which is Olum's endpoint-surface condition~\cite{Olum1998}, with global competitors included.

The stronger assumption matters: Eq.~\eqref{eq:axismaximum} alone permits displaced rays that tie the axis, for example within a finite-width exactly constant transverse core. The strict quartic radial profile used below excludes such ties even though its quadratic radial derivative on the axis vanishes. The negative axial integral in Sec.~\ref{sec:axialbounds}, for a nonconstant matched axial profile, also guarantees nonzero screen curvature somewhere between the endpoints and hence null genericity along the candidate ray. The proof of Eq.~\eqref{eq:Olumcondition} itself does not assume an energy condition or genericity.

\begin{figure*}[tbp]
\includegraphics[width=0.92\textwidth]{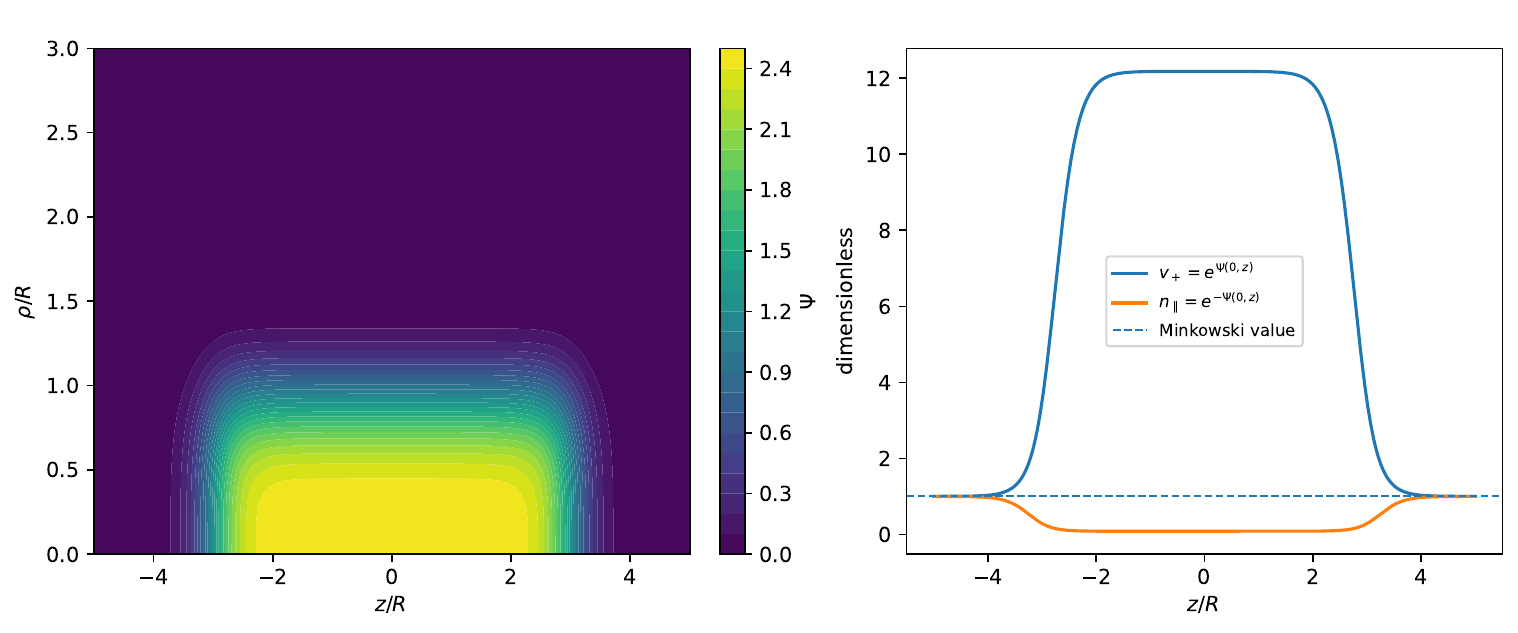}
\caption{Rapidly decaying finite static strain corridor described by Eq.~\eqref{eq:finiteprofile}, with $\Psi_0=2.5$, $D/R=6$, $\delta/R=0.45$, unit lapse, and zero shift. Left: contours and colour scale of the dimensionless strain $\Psi(\varpi,z)$ as a function of $z/R$ and $\varpi/R$. The nearly uniform axial interior is bounded by smooth radial sidewalls and longitudinal endcaps. Right: the axial null coordinate slope $v_+=\ee^{\Psi(0,z)}$ (blue) and optical factor $n_\parallel=\ee^{-\Psi(0,z)}$ (orange), plotted against $z/R$. They are reciprocal, approach the Minkowski value $1$ (dashed line) outside the corridor, and respectively increase and decrease in its interior. The reduced optical factor gives a shorter axial coordinate transit time relative to the asymptotically normalized reference geometry; the coordinate slope is not a locally measured signal speed. These plots use the uncut profile, whose rapidly decaying tails are nonzero. Exact boundary matching instead uses Eq.~\eqref{eq:cutoffprofile}.}
\label{fig:corridor}
\end{figure*}

Figure~\ref{fig:corridor} illustrates a smooth, rapidly decaying profile, rather than a compactly supported one. To distinguish the corridor plateau length from the endpoint separation $L$, denote it by $D$. The example is
\begin{align}
\Psi(\varpi,z)={}&\Psi_0\exp\!\left[-\left(\frac{\varpi}{R}\right)^4\right]
F_D(z),
\label{eq:finiteprofile}\\
F_D(z)={}&\frac12\left[
\tanh\!\left(\frac{z+D/2}{\delta}\right)
-\tanh\!\left(\frac{z-D/2}{\delta}\right)
\right].
\end{align}

The scale $R>0$ controls the transverse width, $\delta>0$ controls the longitudinal transition width, and $D>0$ sets the nominal plateau length. The dimensionless parameter $\Psi_0$ fixes the strain amplitude, and $F_D(z)$ is the longitudinal envelope; for $D\gg\delta$ it is approximately unity over the central plateau. The quartic radial exponent gives a smooth even axis, $\Psi_{,\varpi\varpi}(0,z)=0$, and vanishing quadratic radial variation. For $\Psi_0>0$, $D>0$, and $\delta>0$, $F_D\ge0$, and Eq.~\eqref{eq:axismaximum} holds. The profile has nonzero tails at every finite endpoint, so exact endpoint matching is not a property of Eq.~\eqref{eq:finiteprofile} itself.

An explicit smooth matching prescription is useful. Let $h(s)=\exp(-1/s)$ for $s>0$ and $h(s)=0$ for $s\le0$, and define
\begin{equation}
 \Theta(s)=\frac{h(1-s)}{h(1-s)+h(s)}.
 \label{eq:smoothstep}
\end{equation}
The dimensionless argument $s$ parametrizes a smooth transition: $\Theta=1$ for $s\le0$ and $\Theta=0$ for $s\ge1$. The radii $r_0,r_1$ and longitudinal distances $z_0,z_1$, with $0<r_0<r_1$ and $0<z_0<z_1$, delimit the regions where the strain is unchanged, smoothly suppressed, and set to zero. Let us define
\begin{align}
 \chi_r(\varpi)&=\Theta\!\left(\frac{\varpi^2-r_0^2}{r_1^2-r_0^2}\right),\nonumber\\
 \chi_z(z)&=\Theta\!\left(\frac{z^2-z_0^2}{z_1^2-z_0^2}\right),\nonumber\\
 \Psi_c(\varpi,z)&=\chi_r(\varpi)\chi_z(z)\Psi(\varpi,z).
 \label{eq:cutoffprofile}
\end{align}
The squared arguments ensure smoothness in Cartesian coordinates and reflection symmetry. Endpoints at $|z|>z_1$ have flat neighborhoods; the support is compact. For the positive quartic profile, $0\le\chi_r,\chi_z\le1$ preserve the weak axial maximum, and strict inequality holds for every $\varpi>0$ on any interior interval where $\chi_z>0$. Moreover $a(z)\equiv\partial_\varpi^2\Psi_c(0,z)$, the quadratic radial derivative on the axis, vanishes for this cutoff quartic profile. One concrete choice is $r_0=2R$, $r_1=3R$, $z_0=4R$, $z_1=5R$, with endpoint coordinates $z_A=-6R$, $z_B=6R$. All source and transit quantities for this exactly matched example are evaluated using derivatives of $\Psi_c$, including derivatives of the cutoffs. The uncut numerical values reported below refer instead to Eq.~\eqref{eq:finiteprofile}.

For the uncut illustrative profile, define an asymptotic reference advance by the convergent limit
\begin{equation}
\Delta T_\infty=\int_{-\infty}^{\infty}\left[1-\ee^{-\psi(z)}\right]\dd z>0.
\label{eq:asymptoticadvance}
\end{equation}
It is the limit of reference-minus-physical transit times as the endpoints recede into the asymptotic regions. The unit lapse still makes $t$ proper time for static clocks at finite endpoints, but exact local metric matching requires the cutoff construction.

\subsection{Timelike stationary Killing field and two-way local axial propagation}

The physical norm of the shift is
\begin{equation}
\|\bm\beta\|_\gamma^2
=\gamma_{ij}\beta^i\beta^j
=\ee^{-2\Psi}V^2.
\label{eq:shiftnorm}
\end{equation}
For a stationary configuration, the Killing vector \(\partial_t\) is timelike precisely when
\begin{equation}
|V|<\alpha\ee^{\Psi}.
\label{eq:subcrit}
\end{equation}
In the same regime, Eq.~\eqref{eq:nullspeed} gives
\begin{equation}
V-\alpha\ee^{\Psi}<0<V+\alpha\ee^{\Psi},
\end{equation}
so the two local axial null directions have opposite coordinate signs.  Hence a large boundary-matched time advance need not require \(\|\bm\beta\|_\gamma>\alpha\).  This is a local kinematic statement; it is not, by itself, a theorem excluding a global event horizon in a more general stationary or time-dependent extension.

\subsection{Optical interpretation}

For the static unit-lapse case, null rays extremize the optical length of the Riemannian metric
\begin{equation}
\dd\sigma_{\rm opt}^2
=\ee^{\Psi}(\dd\varpi^2+\varpi^2\dd\phi^2)
+\ee^{-2\Psi}\dd z^2.
\label{eq:opticalmetric}
\end{equation}
The principal optical indices may therefore be read as
\begin{equation}
n_\perp=\ee^{\Psi/2},
\qquad
n_\parallel=\ee^{-\Psi}.
\label{eq:indices}
\end{equation}
The corridor behaves geometrically like a strongly anisotropic optical medium: longitudinal optical distance is reduced while transverse optical distance is increased.  Transformation optics uses the converse mapping, designing media whose rays mimic prescribed curved geometries \cite{LeonhardtPhilbin2008,CrudoOBrien2009}.  Here the optical language is only an interpretation of an actual spacetime metric; it does not supply the gravitational source.

The determinant-preserving relation
\begin{equation}
n_\parallel n_\perp^2=1
\end{equation}
encodes the same anisotropic compensation as Eq.~\eqref{eq:detgamma}.  This connection may nevertheless be useful for analogue experiments because it separates the kinematic ray geometry from the much harder problem of generating the corresponding gravitational stress-energy.

The indices in Eq.~\eqref{eq:indices} are directional optical-length factors relative to the reference coordinates. A local orthonormal observer still measures vacuum light speed equal to one. An analogue medium can reproduce selected ray trajectories, subject to its constitutive and dispersive limitations; it does not reproduce the gravitational source or establish a faster-than-light front velocity in the laboratory.

\section{Exact sources and energy-condition obstructions}
\label{sec:sources}

\subsection{Static curvature and the full source}

We now set \(\alpha=1\) and consider a stationary axisymmetric strain \(\Psi=\Psi(\varpi,z)\).  A direct calculation from Eq.~\eqref{eq:spatialmetric} gives
\begin{align}
\cR={}&\ee^{-\Psi}\Bigg[
\Psi_{,\varpi\varpi}+\frac{\Psi_{,\varpi}}{\varpi}
-2\Psi_{,\varpi}^2
\nonumber\\
&\hspace{1.5cm}-2\ee^{3\Psi}\Psi_{,zz}
-\frac{7}{2}\ee^{3\Psi}\Psi_{,z}^2
\Bigg].
\label{eq:R3general}
\end{align}
The expression is finite on the axis under Eq.~\eqref{eq:axisreg}; the ratio \(\Psi_{,\varpi}/\varpi\) approaches \(\Psi_{,\varpi\varpi}\) there for a smooth even profile.

The derivations and verification scripts were developed with assistance from OpenAI ChatGPT (GPT-6). The accompanying reproducibility scripts independently construct the spatial Christoffel symbols and Ricci tensor, check the tube pressures, finite transverse trace, axial contraction and stationary extrinsic-curvature identity, and evaluate the displayed integrals. The numerical quadratures use dimensionless coordinates $\varpi/R,z/R$, analytic profile derivatives, and independent integrations of both sides of the energy balance.

For a static corridor, \(V=0\), one has \(K_{ij}=0\).  The Hamiltonian constraint
\begin{equation}
16\pi\eps_n=\cR+K^2-K_{ij}K^{ij}
\label{eq:Hconstraint}
\end{equation}

Here $K=\gamma^{ij}K_{ij}$ is the trace of the extrinsic curvature, $\cR$ is the scalar curvature of the spatial metric, and $\eps_n$ denotes energy density measured by an Eulerian observer, whose four-velocity is the future-directed unit normal $n$.
therefore reduces to
\begin{equation}
16\pi\eps_n=\cR,
\label{eq:staticenergy}
\end{equation}
where \(\eps_n=T_{\mu\nu}n^\mu n^\nu\) is the Eulerian energy density.

The complete ultrastatic source follows from $R_{tt}=R_{ti}=0$ and $R_{ij}={}^{(3)}R_{ij}$:
\begin{align}
8\pi\eps_n&=\tfrac12\cR,\qquad j_i=0,\nonumber\\
8\pi S_{ij}&={}^{(3)}R_{ij}-\tfrac12\cR\,\gamma_{ij}.
\label{eq:staticfullsource}
\end{align}
Here $T_{\mu\nu}$ is the required stress-energy tensor, $j_i=-T_{\mu\nu}n^\mu\gamma^\nu{}_i$ is the Eulerian momentum density, and $S_{ij}=\gamma^\mu{}_i\gamma^\nu{}_jT_{\mu\nu}$ is the spatial stress tensor; $\gamma^\mu{}_i$ denotes projection onto the slice. The principal pressures $p_a$ are the eigenvalues of this spatial stress in an orthonormal frame. Hence $\eps_n+\sum_a p_a=0$ in the principal-stress frame. 
For any unit spatial vector $s^i$, satisfying $\gamma_{ij}s^is^j=1$, the vector $k=n+s$ is future-directed and null, with normalization $-n\cdot k=1$. the Einstein equation gives
\begin{equation}
8\pi T_{\mu\nu}k^\mu k^\nu={}^{(3)}R_{ij}s^is^j.
\label{eq:staticNEC}
\end{equation}
The NEC therefore requires the entire spatial Ricci tensor to be positive semidefinite, not just a nonnegative scalar curvature. The metric determines a conserved effective tensor $T_{\mu\nu}=G_{\mu\nu}/(8\pi)$ through the contracted Bianchi identity; it does not determine a matter action or its perturbation equations.

\subsection{Finite-profile Ricci spectrum and null directions}
\label{sec:Riccispectrum}

Use the orthonormal basis $e_{\hat\varpi}=e^{-\Psi/2}\partial_\varpi$, $e_{\hat\phi}=e^{-\Psi/2}\varpi^{-1}\partial_\phi$, $e_{\hat z}=e^\Psi\partial_z$. Away from the axis, let $A={}^{(3)}R_{\hat\varpi\hat\varpi}$, $B={}^{(3)}R_{\hat\phi\hat\phi}$, and $C={}^{(3)}R_{\hat z\hat z}$ denote the diagonal spatial Ricci components, and $H={}^{(3)}R_{\hat\varpi\hat z}$ the radial--longitudinal coupling. These curvature symbols are unrelated to the endpoint worldlines or the mixed deformation tensor $C^i{}_j$. Explicitly,
\begin{align}
 A={}&\tfrac12e^{-\Psi}\left(\Psi_{,\varpi\varpi}-\frac{\Psi_{,\varpi}}\varpi-3\Psi_{,\varpi}^2\right)\nonumber\\
 &-\tfrac12e^{2\Psi}\left(\Psi_{,zz}+2\Psi_{,z}^2\right),\nonumber\\
 B={}&\tfrac12e^{-\Psi}\left(-\Psi_{,\varpi\varpi}+\frac{\Psi_{,\varpi}}\varpi+\Psi_{,\varpi}^2\right)\nonumber\\
 &-\tfrac12e^{2\Psi}\left(\Psi_{,zz}+2\Psi_{,z}^2\right),\nonumber\\
 C={}&e^{-\Psi}\left(\Psi_{,\varpi\varpi}+\frac{\Psi_{,\varpi}}\varpi-\Psi_{,\varpi}^2\right)\nonumber\\
 &-e^{2\Psi}\left(\Psi_{,zz}+\tfrac32\Psi_{,z}^2\right),\nonumber\\
 H={}&-\tfrac12e^{\Psi/2}\left(\Psi_{,\varpi z}+\Psi_{,\varpi}\Psi_{,z}\right).
 \label{eq:RicciABCH}
\end{align}
Their smooth-axis limits are understood in a Cartesian frame. The complete Ricci matrix and spectrum are
\begin{align}
 {}^{(3)}R_{\hat a\hat b}&=
 \begin{pmatrix}A&0&H\\0&B&0\\H&0&C\end{pmatrix},\nonumber\\
 \lambda_\phi&=B,\quad
 \lambda_\pm=\frac{A+C\pm\sqrt{(A-C)^2+4H^2}}2.
 \label{eq:Riccispectrum}
\end{align}
Their trace reproduces Eq.~\eqref{eq:R3general}. Since $j_i=0$ and the spatial stress is symmetric, the ultrastatic stress-energy tensor is of Hawking--Ellis type I: it has a timelike eigenvector and three orthogonal spatial principal directions. Its principal pressures are $p_a=(\lambda_a-\cR/2)/(8\pi)$. Thus the NEC is equivalent to
\begin{equation}
 B\ge0,\quad A\ge0,\quad C\ge0,\quad AC-H^2\ge0.
 \label{eq:finiteNECtest}
\end{equation}
Here the conditions on the $2\times2$ block are necessary and sufficient for positive semidefiniteness. The WEC follows as well because $\eps_n=(A+B+C)/(16\pi)\ge0$ whenever the NEC holds in this ultrastatic family. This implication is special to the source~\eqref{eq:staticfullsource}, not a general equivalence of energy conditions. The dominant energy condition (DEC) requires nonnegative energy density and causal energy flux for every timelike observer; in the principal frame it is $\eps_n\ge|p_a|$. For this source it requires $\lambda_a\le\cR$ as well as $\lambda_a\ge0$. The upper inequalities follow automatically when all eigenvalues are nonnegative, since $\cR=\sum_a\lambda_a$. Hence the NEC, WEC and DEC are equivalent for this particular ultrastatic effective source; this equivalence does not extend to general lapse or shifted geometries.

With the normalization $k=n+s$ fixed as in Eq.~\eqref{eq:staticNEC}, the minimum of $T_{\mu\nu}k^\mu k^\nu$ over unit spatial directions $s$ is $\min(B,\lambda_-)/(8\pi)$. If $\lambda_-$ is the minimum and $H\ne0$, a corresponding meridional eigenvector is proportional to $(H,0,\lambda_--A)$; if $B$ is the minimum, the azimuthal null direction realizes it. Degeneracies permit multiple minimizing directions. This removes the need to infer the full NEC from a selected observer or one diagonal component. Figure~\ref{fig:diagnostics} maps the smallest eigenvalue for the finite example.

\subsection{Finite-corridor energy balance and zero-mass rigidity}

The finite-corridor profile has a useful exact integrated identity.  Assume \(\Psi\) and its first derivatives decay fast enough at large \(\varpi\) and \(|z|\), and impose regularity at \(\varpi=0\).  Since \(\sqrt\gamma=\varpi\), integration by parts in \(\varpi\) and \(z\) gives
\begin{align}
&\int_0^\infty\!\dd\varpi
\int_{-\infty}^{\infty}\!\dd z\,\varpi\cR
\nonumber\\
&\qquad=\int_0^\infty\!\dd\varpi
\int_{-\infty}^{\infty}\!\dd z\,\varpi
\left[-\ee^{-\Psi}\Psi_{,\varpi}^2
+\frac12\ee^{2\Psi}\Psi_{,z}^2\right].
\label{eq:balanceidentity}
\end{align}
Consequently the Eulerian matter-energy volume integral of the static axisymmetric geometry is
\begin{align}
\mathcal E_n={}&\int_{\Sigma_t}\eps_n\sqrt\gamma\,\dd^3x
\nonumber\\
={}&\frac18\int_0^\infty\!\dd\varpi\int_{-\infty}^{\infty}\!\dd z\,\varpi
\left[-\ee^{-\Psi}\Psi_{,\varpi}^2
+\frac12\ee^{2\Psi}\Psi_{,z}^2\right].
\label{eq:Enfinite}
\end{align}

The hypersurface $\Sigma_t$ is the complete constant-$t$ spatial slice, and $\sqrt\gamma\,\dd^3x$ is its proper volume element. Thus $\mathcal E_n$ is a signed matter-energy integral measured in the Eulerian frame, not an integral of the absolute amount of negative energy.
Equation~\eqref{eq:Enfinite} is one of the central source identities of this paper.  The radial-gradient sector is manifestly negative after integration, whereas the longitudinal-gradient sector is manifestly positive.  For long corridor profiles these sectors are concentrated predominantly near the sidewalls and endcaps, respectively, but that spatial language is only an interpretation of the exact gradient decomposition.  Thus a finite corridor is not governed by a universal negative sign for \(\mathcal E_n\).

The ADM mass is an asymptotic surface charge determined by the spatial metric at infinity. The quantity $\mathcal E_n$ is a different measure. It is the volume integral of the matter energy density measured by the Eulerian observers. A sufficient condition for zero ADM mass is exact Euclidean geometry outside a compact set, or Cartesian falloff $\partial^m\Psi=O(\mathsf r^{-1-\eta-m})$ for $m=0,1,2$ and $\eta>0$, with $\mathsf r=\sqrt{x^2+y^2+z^2}$. These conditions also make the scalar curvature absolutely integrable. The matter integral $\mathcal E_n$ can nevertheless be nonzero. This distinction is essential when discussing the ``energy cost'' of a compact shortcut.

There is a stronger pointwise obstruction for complete, nonflat static corridors on $\mathbb R^3$ that are exactly Euclidean outside a compact region, or decay sufficiently rapidly to have zero ADM mass. If $\eps_n\ge0$ everywhere, Eq.~\eqref{eq:staticenergy} gives $\cR\ge0$. The Riemannian positive-mass theorem and its zero-mass rigidity then imply that the spatial metric is isometric to Euclidean space~\cite{SchoenYau1979}. A nonflat member must consequently have $\eps_n<0$ somewhere, violating the WEC. This conclusion uses completeness, the stated asymptotic decay, and nonflatness; it does not follow from the sign of $\mathcal E_n$ alone, and it is not an NEC theorem by itself.

\begin{figure*}[tbp]
\includegraphics[width=0.92\textwidth]{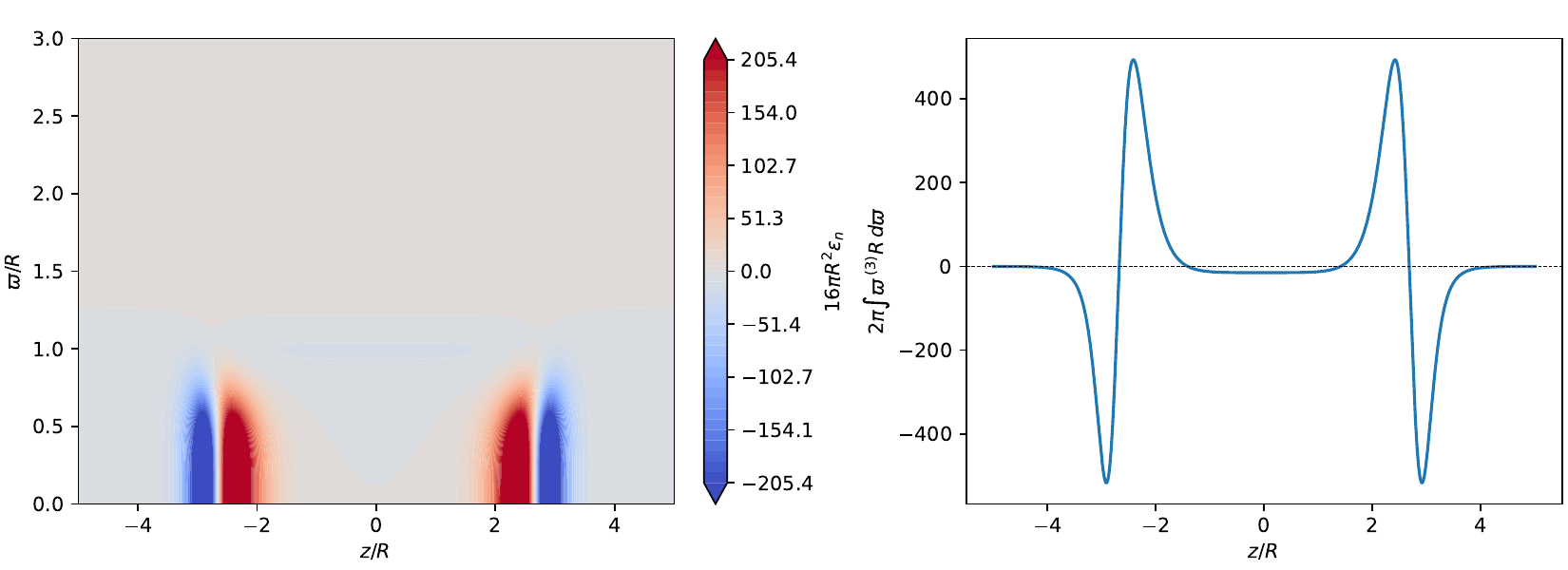}
\caption{Source structure of the uncut finite static corridor in Fig.~\ref{fig:corridor}, with the same parameters and normalization. Left: the dimensionless Eulerian density $16\pi R^2\eps_n=R^2\cR$ as a function of $z/R$ and $\varpi/R$. Red and blue denote positive and negative density, respectively; the colour scale quantifies their magnitude. The strongest contributions lie near the longitudinal endcaps, with weaker structure along the radial transition. Right: the transverse curvature integral $2\pi\int_0^\infty\varpi\cR(\varpi,z)\,\dd\varpi$ as a function of $z/R$; the dotted horizontal line marks zero. The positive and negative lobes near each endcap, together with the negative interior contribution, display the competition between radial and longitudinal gradients discussed in Eq.~\eqref{eq:balanceidentity}. Integration of this curve over $z/R$ gives $16\pi\mathcal E_n/R$, the signed Eulerian matter-energy volume integral in this normalization. Its sign does not by itself determine the local null energy condition or the ADM mass.}
\label{fig:energy}
\end{figure*}

Figure~\ref{fig:energy} displays the local Hamiltonian density and its transverse integral for the uncut decaying profile in Fig.~\ref{fig:corridor}. Direct quadrature gives the dimensionless balance
\begin{equation}
\frac{\mathcal E_n}{R}\simeq-1.73476+2.49603=0.76127,
\label{eq:numericalbalance}
\end{equation}
where the first and second contributions are the radial and longitudinal terms in Eq.~\eqref{eq:Enfinite}. Thus this example has positive integrated Eulerian energy while retaining local negative density. The integral is a geometrized energy in units $G=c=1$; the corresponding SI energy is $(c^4/G)\mathcal E_n$.

For reproducibility, integration on $0\le\varpi/R\le3$, $|z|/R\le9$ with composite Simpson quadrature gives $\mathcal E_n/R=0.7612689113$ for radial grids of 401, 801 and 1601 points (and longitudinal grids of $3N$ points). Direct curvature integration differs from the gradient balance by less than $4\times10^{-10}R$ already on the coarsest grid. Enlarging the integration domain tests the neglected rapidly decaying tails. These convergence checks concern the uncut example; the supplied script evaluates the cutoff model separately.

\subsection{A direct NEC obstruction for finite corridors}
\label{sec:finiteNEC}

Equations~\eqref{eq:RicciABCH} imply the exact transverse trace identity
\begin{equation}
 A+B=-e^{-\Psi}\Psi_{,\varpi}^2-\tfrac12\partial_z^2(e^{2\Psi}).
 \label{eq:transverseTrace}
\end{equation}
Assume, at a fixed $\varpi>0$, that $\partial_z(e^{2\Psi})\to0$ at both longitudinal ends, and that the terms below are integrable. Then
\begin{equation}
 \int_{-\infty}^\infty(A+B)\,\dd z
 =-\int_{-\infty}^\infty e^{-\Psi}\Psi_{,\varpi}^2\,\dd z.
 \label{eq:finiteNECintegral}
\end{equation}
If the radial gradient is nonzero on a set of positive longitudinal measure, this integral is strictly negative. Somewhere at that radius, $A+B<0$, so at least one of $T(n+e_{\hat\varpi},n+e_{\hat\varpi})$ and $T(n+e_{\hat\phi},n+e_{\hat\phi})$ is negative by Eq.~\eqref{eq:staticNEC}. Thus every nontrivial smooth transversely localized finite corridor with these longitudinal decay properties violates the NEC somewhere. Indeed, transverse localization and nontriviality imply a nonzero radial derivative at some point, and smoothness extends it to a longitudinal interval. This result needs neither an axial maximum nor zero-mass rigidity.

The integral~\eqref{eq:finiteNECintegral} uses the coordinate $z$ to establish a local violation. It is not an averaged null energy condition (ANEC) integral, which integrates $T_{\mu\nu}k^\mu k^\nu$ along an affinely parametrized null geodesic: curves generated by these fixed-radius null directions need not be geodesic, and $z$ is not their affine parameter. It complements the invariant complete-ray calculation below. Quantifiers based on volume integration and averaged null energy serve different purposes~\cite{VisserKarDadhich2003,LoboEnergy2005}; neither should be replaced by the sign of $\mathcal E_n$.

\subsection{Infinite-tube stresses and strict exoticity}

The translationally invariant limit \(\Psi=\Psi(\varpi)\) is not asymptotically flat along \(z\), but it is analytically valuable because it isolates the sidewall physics.  Equation~\eqref{eq:R3general} reduces to
\begin{equation}
\cR=\ee^{-\Psi}\left(
\Psi''+\frac{\Psi'}{\varpi}-2\Psi'^2
\right),
\label{eq:R3tube}
\end{equation}
where primes denote \(\dd/\dd\varpi\).

For \(V=0\), the four-dimensional metric is ultrastatic.  In an orthonormal frame, Einstein's equations give
\begin{align}
8\pi\eps_n&=\frac{\ee^{-\Psi}}{2}
\left(\Psi''+\frac{\Psi'}{\varpi}-2\Psi'^2\right),
\label{eq:rhoTube}\\
8\pi p_{\varpi}&=-\frac{\ee^{-\Psi}}{2}
\left(\Psi'^2+\frac{2\Psi'}{\varpi}\right),
\label{eq:prTube}\\
8\pi p_{\phi}&=\frac{\ee^{-\Psi}}{2}
\left(3\Psi'^2-2\Psi''\right),
\label{eq:pphiTube}\\
8\pi p_z&=\frac{\ee^{-\Psi}}{2}
\left(\Psi''+\frac{\Psi'}{\varpi}\right).
\label{eq:pzTube}
\end{align}
The principal null-energy contractions are therefore
\begin{align}
8\pi(\eps_n+p_{\varpi})
&=\frac{\ee^{-\Psi}}{2}
\left(\Psi''-\frac{\Psi'}{\varpi}-3\Psi'^2\right),
\label{eq:NECr}\\
8\pi(\eps_n+p_{\phi})
&=\frac{\ee^{-\Psi}}{2}
\left(-\Psi''+\frac{\Psi'}{\varpi}+\Psi'^2\right),
\label{eq:NECphi}\\
8\pi(\eps_n+p_z)
&=\ee^{-\Psi}
\left(\Psi''+\frac{\Psi'}{\varpi}-\Psi'^2\right).
\label{eq:NECz}
\end{align}

The diagonal stress is of type I, so the NEC is equivalent to $\eps_n+p_a\ge0$ for each principal direction, and the WEC additionally requires $\eps_n\ge0$. In particular,
\begin{equation}
8\pi\left[(\eps_n+p_\varpi)+(\eps_n+p_\phi)\right]
=-\ee^{-\Psi}\Psi'^2.
\label{eq:tubeNECobstruction}
\end{equation}
At every point with $\Psi'\ne0$, at least one of these two null contractions is negative. Transverse localization therefore gives a pointwise NEC obstruction, independently of the volume-integral result below.

The Eulerian matter-energy volume integral also has a definite sign in this translationally invariant limit. Let $L_z$ be a chosen coordinate length along the tube; the ratio $\mathcal E_n/L_z$ is independent of that choice. If \(\Psi'(0)=0\), \(\Psi(\infty)=0\), and \(\varpi\Psi'\to0\) asymptotically, with convergent integrals below, then
\begin{align}
\frac{\mathcal E_n}{L_z}
&=2\pi\int_0^\infty \eps_n\,\varpi\,\dd\varpi
\nonumber\\
&=-\frac18\int_0^\infty
\varpi\ee^{-\Psi}\Psi'^2\,\dd\varpi
\le0.
\label{eq:strictnegative}
\end{align}
Equality holds only when \(\Psi'=0\) everywhere, which together with the boundary condition implies \(\Psi=0\).  Hence every nontrivial, regular, localized static strain tube has strictly negative Eulerian matter-energy volume integral per unit coordinate length.

\begin{figure}[!htbp]
\includegraphics[width=\columnwidth]{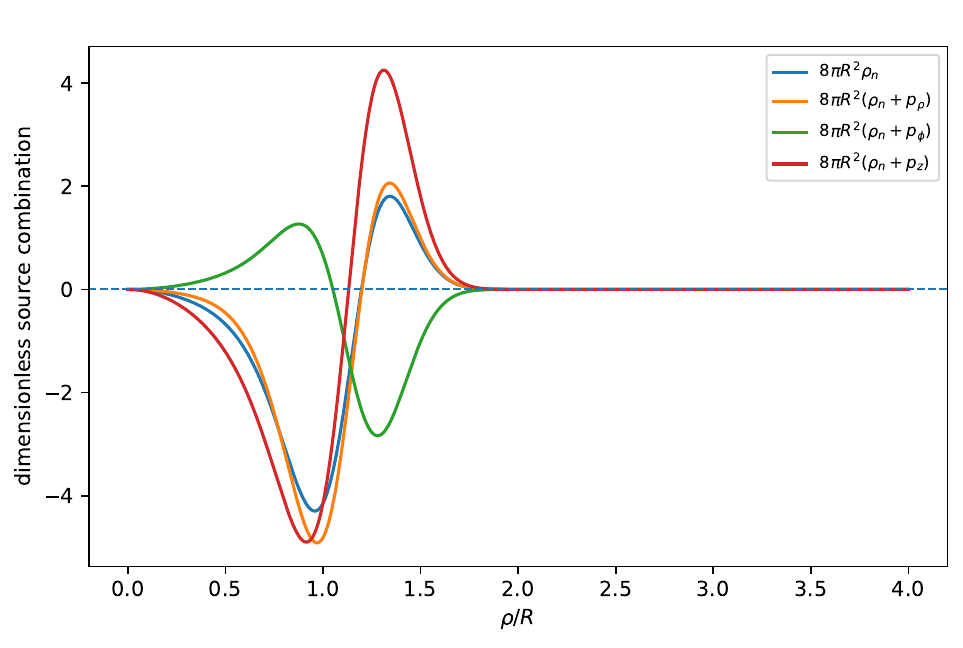}
\caption{Local source combinations for an infinite static strain tube with $\Psi=2\exp[-(\varpi/R)^4]$, unit lapse, and zero shift. The horizontal axis is $\varpi/R$; the vertical axis shows the dimensionless density $8\pi R^2\eps_n$ (blue) and principal null contractions $8\pi R^2(\eps_n+p_{\varpi})$ (orange), $8\pi R^2(\eps_n+p_{\phi})$ (green), and $8\pi R^2(\eps_n+p_z)$ (red), where the pressures are measured in the orthonormal frame. The dashed horizontal line marks zero. The combinations vanish on the quartic-profile axis, acquire different signs through the radial transition, and decay to zero outside the tube. A negative principal null contraction identifies a direction that violates the null energy condition; satisfying that condition requires all three contractions to be nonnegative at the point considered. The density alone therefore does not diagnose the null energy condition, even where it is positive.}
\label{fig:nec}
\end{figure}

This is a direct analogue of the negative-energy obstructions encountered in other spacetime shortcuts, but the origin is geometrically transparent: transverse localization of an axial distance contraction necessarily produces the negative square in Eq.~\eqref{eq:strictnegative}.  Figure~\ref{fig:nec} illustrates the local source combinations for a smooth example \(\Psi=\Psi_0\exp[-(\varpi/R)^4]\).

\subsection{Stationary unit-lapse shifts}

We now impose $\alpha=1$ explicitly and retain stationary $\Psi(\varpi,z)$ and $V(\varpi,z)$. Since $\partial_t\gamma_{ij}=0$, Eq.~\eqref{eq:Kconv} gives
\begin{align}
K_{\varpi\varpi}&=-\frac12 V\ee^{\Psi}\Psi_{,z},
\qquad
K_{\phi\phi}=-\frac12\varpi^2V\ee^{\Psi}\Psi_{,z},
\label{eq:K1}\\
K_{\varpi z}&=-\frac12\ee^{-2\Psi}V_{,\varpi},
\label{eq:K2}\\
K_{zz}&=\ee^{-2\Psi}\left(V\Psi_{,z}-V_{,z}\right).
\label{eq:K3}
\end{align}
The trace is particularly simple, $K=-V_{,z}$.
One finds
\begin{align}
K^2-K_{ij}K^{ij}
={}&-\frac12\ee^{-3\Psi}V_{,\varpi}^2
-\frac32V^2\Psi_{,z}^2
+2V\Psi_{,z}V_{,z}.
\label{eq:Kcombination}
\end{align}

Together with Eq.~\eqref{eq:R3general}, this gives the exact stationary Hamiltonian density for the unit-lapse two-function family. It is not the full stress tensor: momentum and evolution equations are also required to evaluate general shifted energy conditions. For nonunit lapse the displayed $K_{ij}$ acquire a factor $\alpha^{-1}$ and their quadratic combination a factor $\alpha^{-2}$; lapse derivatives additionally enter the stresses.

The simplest shifted control problem is again translationally invariant: \(\Psi=\Psi(\varpi)\), \(V=V(\varpi)\).  Then \(K=0\) and
\begin{equation}
16\pi\eps_n
=\ee^{-\Psi}\left(\Psi''+\frac{\Psi'}{\varpi}-2\Psi'^2\right)
-\frac12\ee^{-3\Psi}V'^2.
\label{eq:rhovshift}
\end{equation}
Hence
\begin{align}
\frac{\mathcal E_n}{L_z}
={}&-\frac18\int_0^\infty
\varpi\ee^{-\Psi}\Psi'^2\,\dd\varpi
\nonumber\\
&-\frac1{16}\int_0^\infty
\varpi\ee^{-3\Psi}V'^2\,\dd\varpi
\le0.
\label{eq:shiftnegative}
\end{align}
The shift contribution is itself negative semidefinite.  Therefore, in this stationary cylindrical subclass, adding an Alcubierre-like axial advection field cannot repair the integrated Eulerian exoticity of the intrinsic strain tube; it makes the bound more negative unless \(V'\equiv0\).

The finite stationary corridor is richer: $-\tfrac32V^2\Psi_{,z}^2$ is nonpositive, whereas $2V\Psi_{,z}V_{,z}$ can have either sign. Longitudinal variation of the shift, endcap curvature, and transverse localization must therefore be retained separately in the source analysis. No time evolution has been included in these stationary identities.

\subsection{Axial averaged null-energy obstruction}

The finite static corridor admits an exact null-source test without solving the complete Ricci eigenvalue problem away from the axis. On the forward axial null geodesic choose an affine parameter $\lambda$ with $\dd t/\dd\lambda=1$, fixing the normalization by the asymptotic unit-lapse clocks. Then
\begin{equation}
k=\partial_t+\ee^{\psi}\partial_z,\qquad
\dd\lambda=\ee^{-\psi}\dd z.
\label{eq:axialaffine}
\end{equation}
Write $a(z)=\Psi_{,\varpi\varpi}(0,z)$. Smooth-axis regularity and Eq.~\eqref{eq:staticNEC} give
\begin{equation}
8\pi T_{\mu\nu}k^\mu k^\nu
=2\ee^{-\psi}a-\ee^{2\psi}\left(\psi''+\tfrac32\psi'^2\right),
\label{eq:axialNECfinite}
\end{equation}
where these primes denote $\dd/\dd z$, rather than the radial derivatives used in the tube formulas. The affine parameter $\lambda$ labels the null geodesic and is distinct from the strain and Ricci eigenvalues. Let us suppose that $\psi$ and $a$ decay sufficiently rapidly for the following integrals to converge and $\ee^\psi\psi'\to0$ at both ends. Integration by parts on the complete ray yields
\begin{align}
\mathcal I_{\rm ax}
&\equiv\int_{-\infty}^{\infty}T_{\mu\nu}k^\mu k^\nu\dd\lambda\nonumber\\
&=\frac{1}{8\pi}\int_{-\infty}^{\infty}
\left[2\ee^{-2\psi}a-\tfrac12\ee^\psi\psi'^2\right]\dd z.
\label{eq:axialANEC}
\end{align}
If the strain is maximal on the axis, Eq.~\eqref{eq:axismaximum} implies $a\le0$. A nonconstant localized axial profile then gives $\mathcal I_{\rm ax}<0$: the averaged null energy condition (ANEC) is violated along a complete achronal null geodesic. This is a direct calculation for the specified ultrastatic family and requires no assumption that Olum's endpoint-surface criterion holds.

For the quartic-core profile in Eq.~\eqref{eq:finiteprofile}, $a=0$ identically. The obstruction resides in its longitudinal transitions:
\begin{equation}
\mathcal I_{\rm ax}=-\frac{1}{16\pi}\int\ee^\psi\psi'^2\dd z<0.
\label{eq:quarticANEC}
\end{equation}
For the parameters of Fig.~\ref{fig:corridor}, $R\mathcal I_{\rm ax}\simeq-0.74918$. Together with Eq.~\eqref{eq:numericalbalance}, this shows explicitly that positive integrated Eulerian energy does not remove a null-energy obstruction. Rescaling the affine parameter changes the magnitude by a positive factor, while the sign is invariant.

\subsection{Quantitative advance--ANEC bounds}
\label{sec:axialbounds}
Consider the complete matched ultrastatic corridor with $\Psi(\varpi,z)=0$ for $z\notin[z_A,z_B]$, $\psi\ge0$, and an axial maximum, so $a\le0$. Keep the affine normalization~\eqref{eq:axialaffine}. Let $y=e^{\psi/2}$ be a dimensionless auxiliary field and retain $L=z_B-z_A$ for the reference endpoint separation. Then $y(z_A)=y(z_B)=1$ and Eq.~\eqref{eq:axialANEC} gives
\begin{equation}
 -\mathcal I_{\rm ax}\ge\frac1{4\pi}\int_{z_A}^{z_B}y'^2\,\dd z.
 \label{eq:DirichletANEC}
\end{equation}
Here contributions outside the endpoint interval are zero. The elementary inequality $1-y^{-2}\le2(y-1)$ yields $\Delta T_{\rm ref}\le2\int(y-1)\dd z$. To estimate this integral, use $x=z-z_A$, $f=y-1$, and $w=x(L-x)/2$, with $f(0)=f(L)=0$. Integration by parts and Cauchy--Schwarz give
\begin{align}
 \left(\int_0^L f\,\dd x\right)^2
 =\left(\int_0^L w'f'\,\dd x\right)^2
 \le\frac{L^3}{12}\int_0^L f'^2\,\dd x.
 \label{eq:advanceSobolev}
\end{align}
Combining these relations proves
\begin{equation}
 -\mathcal I_{\rm ax}\ge
 \frac{3(\Delta T_{\rm ref})^2}{4\pi L^3}.
 \label{eq:advanceANECbound}
\end{equation}
This is a conservative finite-length bound, not a claim of optimality at finite amplitude. Let $\psi_{\max}=\max_{[z_A,z_B]}\psi(z)$ be the peak axial strain. If $y$ reaches $y_m=e^{\psi_{\max}/2}>1$ at an interior point a reference coordinate distance $\ell$ from the left endpoint, two further Cauchy--Schwarz estimates give
\begin{align}
 \int_0^L y'^2\,\dd x
 &\ge(y_m-1)^2\left(\frac1\ell+\frac1{L-\ell}\right)
 \nonumber\\
 &\ge\frac{4(y_m-1)^2}{L},
\end{align}
so
\begin{equation}
 -\mathcal I_{\rm ax}\ge
 \frac{(e^{\psi_{\max}/2}-1)^2}{\pi L}.
 \label{eq:amplitudeANECbound}
\end{equation}
Both bounds have units of inverse length and refer to the same normalized ray. They quantify unavoidable negative averaged null energy for a specified length and advance or amplitude. They are not bounds on ADM mass, on the signed volume integral, or on a universal quantum-field energy functional. For the cutoff quartic model, $a=0$ and the Dirichlet relation~\eqref{eq:DirichletANEC} is an equality; the subsequent finite-amplitude inequalities generally remain strict.

For the cutoff parameters specified after Eq.~\eqref{eq:cutoffprofile}, numerical quadrature gives $\Delta T_{\rm ref}/R=5.982999906$ and $R\mathcal I_{\rm ax}=-0.749181592$; its signed matter integral is $\mathcal E_n/R=0.761273964$. These values include cutoff derivatives. With $L=12R$, both bounds are satisfied. The small change from the uncut example demonstrates convergence of this particular matching prescription, without equating the two geometries.

\subsection{Axial passenger tides}
\label{sec:tides}
\begin{figure*}[tbp]
\centering
\includegraphics[width=0.94\textwidth]{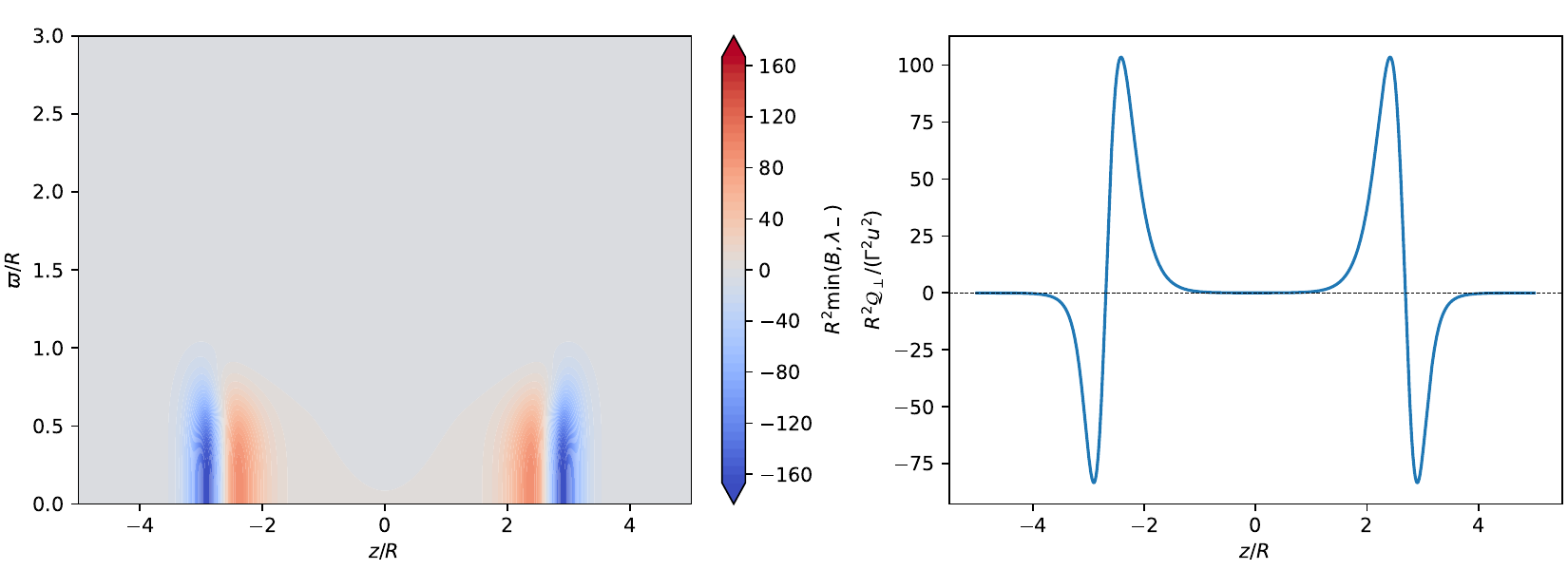}
\caption{Curvature and passenger-tide indicators for the uncut finite static corridor in Fig.~\ref{fig:corridor}, with the same parameters. Left: $R^2\min(B,\lambda_-)$ as a function of $z/R$ and $\varpi/R$, where Eq.~\eqref{eq:Riccispectrum} gives the Ricci eigenvalues and the minimizing spatial direction. This quantity is the minimum null Ricci contraction for $k=n+s$, with $s$ a unit spatial vector, multiplied by $R^2$; equivalently, it is $8\pi R^2\min_s T_{\mu\nu}k^\mu k^\nu$. Blue regions indicate negative values and hence violation of the null energy condition in at least one direction; red regions indicate positive minima. Right: the dimensionless axial transverse tidal factor $R^2\mathscrQ_\perp/(\Gamma^2u^2)=R^2C(0,z)/2$, plotted against $z/R$, with the dotted line marking zero. The two transverse eigenvalues coincide, and the longitudinal eigenvalue vanishes. Positive and negative values correspond to transverse compression and stretching, respectively, under the geodesic-deviation convention in the text. Multiplication by $\Gamma^2u^2/R^2$ gives the physical tidal eigenvalue for a passenger of local axial speed $u$. The normalized factor is interpreted by continuity at $u=0$, while the actual electric tides vanish for a static observer. The endcap signal arises from longitudinal derivatives even though the quadratic radial strain vanishes on the axis.}
\label{fig:diagnostics}
\end{figure*}

We use the Riemann-curvature convention $R^\rho{}_{\sigma\mu\nu}=\partial_\mu\Gamma^\rho_{\nu\sigma}-\partial_\nu\Gamma^\rho_{\mu\sigma}+\Gamma^\rho_{\mu\lambda}\Gamma^\lambda_{\nu\sigma}-\Gamma^\rho_{\nu\lambda}\Gamma^\lambda_{\mu\sigma}$. Here $\Gamma^\rho_{\mu\nu}$ denotes a Christoffel symbol, distinct from the Lorentz factor $\Gamma=(1-u^2)^{-1/2}$. For an axial passenger with four-velocity $U=\Gamma(n+u e_{\hat z})$, let $E_{\hat x},E_{\hat y}$ be orthonormal transverse spatial vectors and $E_{\hat z}=\Gamma(u n+e_{\hat z})$ the longitudinal unit vector in the passenger rest frame. The electric tidal tensor, which governs relative acceleration between neighboring freely falling particles, is defined by $\mathscrQ_{ab}=R_{\mu\alpha\nu\beta}E_a^\mu U^\alpha E_b^\nu U^\beta$. Ultrastaticity eliminates curvature components with time indices, and rotational symmetry gives $R_{\hat x\hat z\hat x\hat z}=R_{\hat y\hat z\hat y\hat z}=C(0,z)/2$. Therefore its passenger-frame eigenvalues are
\begin{align}
 \mathscrQ_\perp={}&\Gamma^2u^2\left[e^{-\psi}a
 -\tfrac12e^{2\psi}\left(\psi''+\tfrac32\psi'^2\right)\right],\nonumber\\
 \mathscrQ_\parallel={}&0,
 \label{eq:passengertides}
\end{align}
with the transverse eigenvalue repeated twice. Geodesic deviation is $\dd^2\xi^a/\dd\tau^2=-\mathscrQ^a{}_b\xi^b$ in a parallel-transported passenger frame. Static observers have zero electric tides, whereas moving passengers experience the boosted spatial curvature. On the quartic axis $a=0$, so the longitudinal transitions supply the entire axial tidal signal. Figure~\ref{fig:diagnostics} displays this factor independently of $\Gamma^2u^2$. A size-dependent traversability constraint is $|\mathscrQ_\perp|\,\ell_{\rm body}\le a_{\rm tol}$ in geometrized units, where $\ell_{\rm body}$ is the transverse separation across the passenger or vehicle and $a_{\rm tol}$ is the tolerated differential acceleration. The separation vector $\xi^a$ describes neighboring geodesics, and $\tau$ is the passenger proper time. This follows the frame-sensitive traversability analysis used for wormholes~\cite{MorrisThorne1988,LoboEnergy2005}, while the curvature here is that of the strained corridor.

\section{Physical discussion and extensions}
\label{sec:physical}
\subsection{Comparison with existing geometries}

The novelty claim requires particular care.  Nonflat spatial slices are not new in warp-drive studies, and neither are volume-changing or even unit-determinant spatial deformations.  Visser and Liberati, for example, discuss a factorized volume-preserving deformation of three-space in the distinct context of unit-lapse Kerr representations \cite{VisserLiberati2022}.  Thus neither \(\det(\bar\gamma^{-1}\gamma)=1\) nor a non-Euclidean intrinsic metric can be claimed as new by itself.

Alcubierre's original construction keeps the spatial slices Euclidean and places the transport kinematics in the shift \cite{Alcubierre1994}.  Nat\'ario imposes vanishing expansion on the warp flow, again with flat intrinsic spatial geometry \cite{Natario2002}.  Van Den Broeck uses spatial geometry to enlarge an internal region while retaining shift-driven transport \cite{VanDenBroeck1999}.  Bobrick and Martire formulate a much broader class of physical warp-drive shells with freedom in spatial capacity and clock rates \cite{BobrickMartire2021}.  Chowdhury studies Alcubierre--Nat\'ario-type transport on nonflat Martel--Poisson slices \cite{Chowdhury2025}.  Barzegar and Buchert emphasize the need to restore spatial curvature, acceleration, vorticity, and general spatial motion \cite{BarzegarBuchert2025}; subsequent work develops broader classifications and curved-space realizations \cite{BarzegarBuchertVigneron2026,BuchertFrackowiak2026}.  Krasnikov's construction is also a close conceptual relative because it treats a prepared route rather than a compact moving bubble \cite{Krasnikov1998,Krasnikov2003}.

The contribution can therefore be stated directly: the specified longitudinal-contraction/transverse-expansion strain produces a reference time advance at $V=0$, a sufficient global-minimality criterion, and exact finite- and infinite-corridor source identities. Equations~\eqref{eq:Enfinite}, \eqref{eq:tubeNECobstruction}, \eqref{eq:shiftnegative}, and \eqref{eq:axialANEC} identify complementary energy obstructions. The construction supplies an explicit sector for comparison and analysis, without requiring an exhaustive historical priority claim.

\subsection{Tensorial strain and curved reference backgrounds}

The scalar ansatz is the simplest member of a more general construction. Let $\bar\gamma_{ij}$ be a reference spatial metric and $S^i{}_j$ a dimensionless strain tensor, distinct from the spatial stress $S_{ij}$ in Eq.~\eqref{eq:staticfullsource}. Require $S^i{}_j$ to be self-adjoint with respect to $\bar\gamma_{ij}$, meaning that $\bar\gamma_{ik}S^k{}_j$ is symmetric, and trace free, $S^i{}_i=0$.

Let us define
\begin{equation}
\gamma_{ij}
=\bar\gamma_{ik}\left(\ee^{2S}\right)^k{}_j.
\label{eq:generalstrain}
\end{equation}

The exponential is a matrix exponential. Self-adjointness makes the resulting spatial metric symmetric and positive definite; trace freedom preserves the reference volume form.
Then
\begin{equation}
\det\gamma
=\det\bar\gamma\,\ee^{2\Tr S}
=\det\bar\gamma.
\label{eq:generaldet}
\end{equation}
Our cylindrical metric corresponds locally to principal strain eigenvalues $\left(\Psi/2,\Psi/2,-\Psi\right)$.
Equation~\eqref{eq:generalstrain} provides a natural language for curved trajectories, rotating principal axes, branching corridors, helical transport channels, and background geometries with nontrivial topology.

A particularly compelling extension is to place the trace-free strain on a wormhole slice. If $s^i$ is a smooth unit vector field selecting the desired transport direction, let $P_{ij}=\bar\gamma_{ij}-s_i s_j$ be the projector onto the two directions transverse to $s^i$. Then
\begin{equation}
\gamma_{ij}^{\rm strained}
=\ee^{-2\Psi}s_i s_j+\ee^{\Psi}P_{ij}
\label{eq:strainedbackground}
\end{equation}
preserves \(\det\bar\gamma\).  
For a spherical two-ended wormhole with reference proper radial coordinate $l\in\mathbb R$, $s=\partial_l$ is a smooth global choice~\cite{MorrisThorne1988}. Topological connectivity, reference-relative radial contraction, and a chosen ADM shift can then be specified separately. The source, regularity, causal structure, and stability must be recalculated for the resulting metric; the cylindrical identities do not transfer unchanged.

In Eq.~\eqref{eq:strainedbackground}, $s_i=\bar\gamma_{ij}s^j$ and $\bar\gamma_{ij}s^is^j=1$. The self-adjoint, trace-free construction ensures a positive spatial metric with the reference volume form, but gives no general shortcut or energy-condition theorem when the principal directions vary. The separation into lapse, shift, and strain is relative to the chosen foliation and reference geometry; it is not a foliation-independent partition of gravitational dynamics.

\subsection{Physical requirements and constrained optimization}

The strain corridor is a metric ansatz, not a formation mechanism.  Einstein's equations tell us the stress-energy tensor required to support the geometry; they do not identify matter capable of producing it.  The exact negative-energy theorem \eqref{eq:strictnegative} for the infinite static tube and the explicit NEC combinations \eqref{eq:NECr}--\eqref{eq:NECz} show that the construction does not evade the familiar exotic-matter problem by semantic redefinition.  Rather, it moves the geometrical origin of the shortcut from shift advection into the intrinsic metric, making a different set of source terms available for analysis.

A useful source optimization should fix an operational transit time and an explicit measure of energy-condition violation. Related distinctions between pointwise, averaged and volume-integrated measures are developed in Refs.~\cite{VisserKarDadhich2003,LoboEnergy2005}; broad discussions of the reverse construction and its physical scope appear in Refs.~\cite{LoboReview2008,AlcubierreLobo2017}. For the stationary unit-lapse family, smooth axisymmetry preserves axial geodesic motion. If $V(0,z)+\ee^{\psi(z)}>0$ along the route, its forward null transit time is
\begin{equation}
\mathcal T[\Psi,V]=\int_{z_A}^{z_B}
\frac{\dd z}{V(0,z)+\ee^{\psi(z)}}.
\label{eq:stationarytime}
\end{equation}
Exact endpoint matching also requires $V=\Psi=0$ near the endpoint worldlines. Equation~\eqref{eq:Taxis} is the zero-shift limit. A constrained problem may then be written as
\begin{equation}
\min_{\Psi,V}\mathfrak E[T_{\mu\nu}],\qquad
L-\mathcal T[\Psi,V]=\Delta T_\star,
\label{eq:optimization}
\end{equation}
Here $\mathfrak E[T_{\mu\nu}]$ is a specified measure of source requirements or energy-condition violation, and $\Delta T_\star>0$ is the prescribed reference time advance. The admissible profiles and curvature bounds must also be specified. If $V$ varies in the optimization, the static time functional cannot be used unchanged. Nor should $\mathcal E_n$ be minimized as a substitute for tests of the NEC and WEC. Numerical curvature and observer-dependent energy-condition tools can support such calculations~\cite{Helmerich2024}.

An operational comparison does not supply the energy or causal process required to create the corridor. It assumes the geometry is already prepared. A formation analysis must specify matter equations, initial data, and switching dynamics; related simulations of warp-drive collapse illustrate the dependence on the chosen dynamical source~\cite{Clough2024}.

A perturbative analysis would need to evolve both the geometry and the supporting matter, rather than perturb the metric while leaving its effective source unspecified. The reverse construction $T_{\mu\nu}=G_{\mu\nu}/(8\pi)$ satisfies Einstein's equations identically but is not a physical equation of state. In particular, conservation of this tensor is insufficient to determine linear or nonlinear stability.

The exact axial tidal tensor~\eqref{eq:passengertides} supplies a practical constraint for the optimization~\eqref{eq:optimization}, together with the advance--ANEC bounds~\eqref{eq:advanceANECbound}--\eqref{eq:amplitudeANECbound}. Off-axis passengers require the full boosted curvature tensor. Smoothness gives finite curvature at each point, while uniform curvature bounds require control of the derivatives as well as strain amplitude. At fixed dimensionless profile, travel-time advances scale linearly with length, curvature as inverse length squared, $\mathcal E_n$ linearly with length, and the normalized ANEC integral as inverse length. Such rescaling does not establish physical realizability.

Quantum restrictions must be applied with their field, state, curvature, and sampling assumptions stated. The thin-wall bounds derived for conventional warp drives~\cite{PfenningFord1997} cannot be imported as numerical constraints on this different geometry. Moreover, a universal lower bound on null-geodesic averages does not exist for the massless minimally coupled scalar field in four-dimensional Minkowski space~\cite{FewsterRoman2003}; that result neither supplies the required curved-space stress tensor nor removes the negative complete-ray integral in Eq.~\eqref{eq:axialANEC}. A semiclassical realization would require an appropriate renormalized tensor and self-consistent backreaction. The horizon-driven instability found for dynamical superluminal warp bubbles~\cite{FinazziLiberatiBarcelo2009} is a warning about those geometries, not a demonstrated instability of this horizonless ultrastatic subclass.

The weak axial maximum establishes global point-to-point minimality. With a strict axial maximum and exactly flat endpoint neighborhoods, Sec.~\ref{sec:Olum} proves the stronger endpoint-surface condition~\cite{Olum1998}. The negative axial integral guarantees genericity somewhere along a nonconstant matched axial ray; no assumption of NEC compliance is made. Gao--Wald time-delay results likewise assume null-energy and null-generic conditions~\cite{GaoWald2000}; the axial ANEC violation derived here prevents treating the present corridor as an energy-respecting counterexample. The exact source obstructions establish limits on this ansatz while leaving its matter completion and dynamical formation open.

\section{Conclusions}\label{sec:conclusions}

The family studied here separates intrinsic spatial strain, defined relative to a fixed reference geometry, from transport generated by an ADM shift. Longitudinal proper lengths contract while transverse proper lengths expand, preserving the reference spatial volume form. In a prepared ultrastatic corridor, this deformation produces an exact travel-time advance between synchronized exterior clocks while preserving ordinary local causal propagation. Smooth bounded strain gives a globally hyperbolic and geodesically complete spacetime. An axial maximum makes the axial light ray globally minimizing; a strict maximum on an interior interval, together with exactly flat endpoint neighborhoods, establishes the stronger endpoint-surface criterion.

The source identities distinguish three quantities with different physical roles: local energy density, the signed Eulerian matter-energy volume integral, and averaged null energy along a complete ray. Every nontrivial localized infinite static tube has a negative matter-energy integral per unit coordinate length, and transverse stationary shift gradients contribute another nonpositive term. A finite corridor instead admits competing radial and longitudinal contributions, so its signed integral can be positive. Under the stated completeness and zero-ADM-mass assumptions, however, a nonflat static corridor must still contain negative local Eulerian density. Independently, the transverse Ricci identity proves local NEC violation for nontrivial transversely localized finite corridors with the specified longitudinal decay. For a nonconstant localized axial-maximum profile, the complete axial ANEC integral is strictly negative. A positive volume-integrated energy therefore does not remove the local or averaged source obstructions.

The quantitative advance--ANEC bounds connect an operational travel-time improvement to the magnitude of the negative complete-ray integral at fixed reference length. The full Ricci spectrum identifies the null directions in which the required source is most restrictive, while the passenger-frame tidal tensor converts curvature into a differential-acceleration constraint. Together these results provide a basis for constrained profile optimization. A natural next step is to determine the sharp finite-amplitude bounds and the profiles that approach them, while imposing smooth exterior matching and uniform curvature and tidal limits. Such a calculation would quantify the tradeoffs between achievable advance, transition widths, and source requirements more precisely than a signed energy integral alone.

A second research direction is to supply a dynamical matter model. One must identify an action or constitutive equations, solve the associated initial-value and constraint equations, and test whether the prescribed stress-energy tensor can be attained. This would connect the prepared geometries analyzed here to questions of formation, switching, and coupled matter--metric stability. In a semiclassical treatment, the corresponding task is to calculate a renormalized stress-energy tensor for specified fields and states and include its backreaction. The present geometric identities constrain these investigations but do not establish that a physical source can form or sustain the corridor.

Finally, the tensorial strain construction offers a route beyond straight axisymmetric corridors. Curved transport directions and wormhole reference slices allow intrinsic distance contraction and global connectivity to be investigated within a common geometric framework. Each extension requires a new analysis of optical minimality, endpoint normalization, regularity, source eigenvalues, and passenger tides. Determining which of the present causal estimates and energy obstructions survive would connect this explicit strain family to a broader theory of spacetime shortcuts, with clearly stated operational advantages and physical limitations.\\

\begin{acknowledgments}
FSNL acknowledges support from the Funda\c{c}\~ao para a Ci\^encia e a Tecnologia (FCT) research grant UID/04434/2025.
\end{acknowledgments}

\end{document}